\pdfoutput=1  
\def\nk{n_\mathrm{K}}
\def\acap{\\ \nonumber \\}

\def\rfr#1{Equation\,(\ref{#1})}
\def\rfrs#1#2{Equations\,(\ref{#1})--(\ref{#2})}

\def\Rfrs#1#2{Equations\,(\ref{#1})--(\ref{#2})}
\def\derp#1#2{\rp{\partial{#1}}{\partial{#2}}}
\def\dert#1#2{\frac{{{\textrm{d}}}{#1}}{{{\textrm{d}}}{#2}}}
\def\virg#1{``#1"}
\def\eqi{\begin{equation}}
\def\eqf{\end{equation}}
\def\rp#1#2{\frac{#1}{#2}}
\def\lb#1{\label{#1}}

\def\ton#1{\left(#1\right)}

\def\grf#1{\left\{#1\right\}}

\RequirePackage[2020-02-02]{latexrelease}
\documentclass{aastex631}
\usepackage{morefloats}
\usepackage[title]{appendix}
\usepackage{textcomp}
\usepackage{booktabs}
\usepackage{multirow}
\usepackage{rotating,tabularx}
\usepackage{float}
\usepackage{enumerate}
\usepackage{rotating}
\usepackage[polutonikogreek,english]{babel}
\usepackage{amsmath,starfont,textgreek,w-greek}
\usepackage[flushleft]{threeparttable}
\usepackage{amsthm}
\usepackage{bigints}
\usepackage{amscd}
\usepackage[mathlines]{lineno}
\usepackage{amssymb,dsfont}
\usepackage{graphicx,epsfig}
\usepackage{txfonts}
\usepackage{xr-hyper}
\usepackage{hyperref}
\usepackage[utf8]{inputenc}
\usepackage{newunicodechar,graphicx}

\usepackage[nointegrals]{wasysym}
\usepackage[caption=false]{subfig}

\allowdisplaybreaks

\makeatletter
\DeclareRobustCommand\ref{%
    \@ifstar\@refstar\T@ref
  }%
  \DeclareRobustCommand\pageref{%
    \@ifstar\@pagerefstar\T@pageref
  }%
 \makeatother

\begin{document}

\title{The no--hair theorems at work in M87$^\ast$}

\shortauthors{L. Iorio}

\author[0000-0003-4949-2694]{Lorenzo Iorio}
\affiliation{Ministero dell' Istruzione e del Merito. Viale Unit\`{a} di Italia 68, I-70125, Bari (BA),
Italy}

\email{lorenzo.iorio@libero.it}

\begin{abstract}
Recently, a perturbative calculation to the first post--Newtonian order has shown that the analytically worked out Lense--Thirring precession of the orbital angular momentum of a test particle following a circular path around a massive spinning primary is able to explain the measured features of the jet precession of the supermassive black hole at the centre of the giant elliptical galaxy M87. It is shown that also the hole's mass quadrupole moment $Q_2$, as given by the no--hair theorems, has a dynamical effect which cannot be neglected, as, instead, done so far in the literature. New allowed regions for the hole's dimensionless spin parameter $a^\ast$ and the effective radius $r_0$ of the accretion disk, assumed tightly coupled with the jet,  are obtained by including both the Lense--Thirring and the quadrupole effects in the dynamics of the effective test particle modeling the accretion disk. One obtains that, by numerically integrating the resulting averaged equations for the rates of change of the angles $\eta$ and $\phi$ characterizing the orientation of the orbital angular momentum with $a^\ast = +0.98$ and $r_0=14.1$ gravitational radii, it is possible to reproduce, both quantitatively and qualitatively, the time series for them recently measured with the Very Long Baseline Interferometry technique. Instead, the resulting time series produced with $a^\ast = -0.95$ and $r_0=16$ gravitational radii turn out to be out of phase with respect to the observationally determined  ones, while maintaining the same amplitudes.
\end{abstract}

%
%
%
\section{Introduction}
It has recently been proven \citep{2024arXiv241108686I} that a simple perturbative calculation to the first post--Newtonian (1pN) order of the Lense--Thirring (LT) effect on the orbit of a test particle moving about a massive spinning object is able to reproduce, both qualitatively and quantitatively, several features of the jet precession in the supermassive black hole (SMBH) M87$^\ast$ recently measured with the Very Long Baseline Interferometry (VLBI) technique \citep{2023Natur.621..711C}. Aim of this paper is to show that, actually, also the mass quadrupole moment\footnote{The possible impact of electric charge was recently investigated \citep{2024arXiv241107481M} in the framework of the Kerr--Newman metric \citep{1965JMP.....6..915N,1965JMP.....6..918N} as well.} of  M87$^\ast$, neglected so far in the literature,
should be fully taken into account in the dynamics of the accretion disk, assumed tightly coupled with the jet \citep{2013Sci...339...49M,2018MNRAS.474L..81L,2024arXiv241010965C}. The imprints of the precession of the latter on the SMBH-accretion disk system was recently investigated in \citep{2024arXiv241010965C}. The introduction of the hole's quadrupole does not spoil the agreement between the perturbative analytical calculation and the characteristics measured in \citep{2023Natur.621..711C}, yielding to different allowed regions for the hole's spin parameter and the radius of the effective circular orbit modeling the precessing disk. As a result, the celebrated \virg{no--hair} (NH) theorems \citep{1967PhRv..164.1776I,1971PhRvL..26..331C,1975PhRvL..34..905R} receive a strong support.

According to the latter ones,
the mass and the spin moments  $\mathcal{M}^\ell$  and  $\mathcal{J}^\ell$  of degree $\ell$ of a rotating BH \citep{1970Natur.226...64B}, whose external spacetime is described by the Kerr metric \citep{1963PhRvL..11..237K,2015CQGra..32l4006T},  are connected by the relation
\eqi
\mathcal{M}^\ell + i\,\mathcal{J}^\ell= M \ton{i\,\rp{J}{cM}}^\ell,\lb{nohair}
\eqf
where $i:=\sqrt{-1}$ is the imaginary unit, $c$ is the speed of light, and $M$ and $J$ are the hole's mass and spin angular momentum, respectively. From \rfr{nohair}, it turns out that the odd mass moments and even spin moments
are identically zero. In particular, the hole's mass moment $\mathcal{M}^0$ of degree $\ell=0$  is  its mass, while its spin dipole moment $\mathcal{J}$ of degree $\ell=1$  is its spin angular momentum. For  a Kerr BH, it is \citep{1986bhwd.book.....S}
\eqi
J =\chi \rp{M^2 G}{c},\,\left|\chi\right|\leq 1,\lb{Jei}
\eqf
where $G$ is the Newtonian constant of gravitation.
Furthermore, the mass quadrupole moment $\mathcal{M}^2$ of degree $\ell=2$, renamed $Q_2$,   is
\eqi
Q_2 = -\rp{J^2}{c^2 M}.\lb{Q2}
\eqf
If $\left|\chi\right|>1$, a naked singularity \citep{1973CMaPh..34..135Y,1991PhRvL..66..994S} without a horizon would occur, implying the possibility of causality violations because of closed timelike curves. Although not yet proven, the cosmic censorship conjecture \citep{1999JApA...20..233P,2002GReGr..34.1141P} states that naked singularities may not be formed via the gravitational collapse of a material body. The dimensionless parameter $\chi$ can also be viewed as the second characteristic length\footnote{Sometimes, the symbol $a$ is used  for $J/\ton{cM}$ itself, in which case it is dimensionally a length. } $J/\ton{cM}$ occurring in the Kerr metric \citep{1963PhRvL..11..237K,2015CQGra..32l4006T} measured in units of the gravitational radius $R_\mathrm{g}:=GM/c^2$. In BH studies, $\chi$ is often denoted with $a$, which is the same symbol usually adopted in celestial mechanics and astrodynamics to denote the semimajor axis of a generally elliptical orbit of a test particle.

The organization of the paper is as follows.
Section \ref{sect:model} offers a review of the perturbative analytical model of the LT and mass quadrupole effects, expressed in terms of the both the Keplerian orbital elements and in vectorial form, for a generic orientation of the primary's spin axis in space. In Section \ref{sect:application}, the previous results  are successfully applied to the measured precession of the jet of M87$^\ast$, assumed strongly coupled with that of the accretion disk. Section \ref{fine} summarizes the findings and offers conclusions.
\section{The analytical model for the orbital precessions}\lb{sect:model}
The inclination $I$ and the longitude of the ascending node $\mathit{\Omega}$ define the orientation of the orbital plane in space. Their LT rates of change averaged over one orbital revolution and valid for an arbitrary orientation of the primary's spin axis $\boldsymbol{\hat{k}}$ turn out to be
\begin{align}
\dot I^\mathrm{LT} \lb{dotILT}& = \rp{2GJ\ton{\boldsymbol{\hat{k}}\boldsymbol\cdot\boldsymbol{\hat{l}}}}{c^2 a^3\ton{1 - e^2}^{3/2}},\acap
\dot{\mathit{\Omega}}^\mathrm{LT} \lb{dotOLT}& = \rp{2GJ\csc I\ton{\boldsymbol{\hat{k}}\boldsymbol\cdot\boldsymbol{\hat{m}}}}{c^2 a^3\ton{1 - e^2}^{3/2}},
\end{align}
where, for a Kerr BH, $J$ is given by \rfr{Jei}. Furthermore, $a$ and $e$ are the orbit's semimajor axis and eccentricity, respectively.
The unit vectors $\boldsymbol{\hat{l}}$ and $\boldsymbol{\hat{m}}$, which lie in the orbital plane and are mutually orthogonal, are defined as \citep{Sof89,1991ercm.book.....B,SoffelHan19,2024gpno.book.....I}
\begin{align}
\boldsymbol{\hat{l}} \lb{elle}&:=\grf{\cos\mathit{\Omega},\sin\mathit{\Omega},0},\acap
\boldsymbol{\hat{m}} \lb{emme} & := \grf{-\cos I \sin\mathit{\Omega},
\cos I \cos\Omega, \sin I}.
\end{align}
\Rfrs{dotILT}{dotOLT} were worked out with a perturbative calculation to the 1pN order by means of the equations for the rates of change of the Keplerian orbital elements in Gaussian or Lagrange form \citep{Sof89,1991ercm.book.....B,2000ssd..book.....M,2003ASSL..293.....B,2005ormo.book.....R,2011rcms.book.....K,2014grav.book.....P,SoffelHan19,2024gpno.book.....I}.

The mean rates of changes of the same orbital elements caused by the oblateness of a massive primary, valid for a generic orientation of $\boldsymbol{\hat{k}}$, can be calculated with the same approach obtaining \citep{2024gpno.book.....I}
\begin{align}
\dot I^{Q_2} \lb{dotIQ2}& = -\rp{3}{2}\nk J_2\ton{\rp{R}{p}}^2\ton{\boldsymbol{\hat{k}}\boldsymbol\cdot\boldsymbol{\hat{h}}}\ton{\boldsymbol{\hat{k}}\boldsymbol\cdot\boldsymbol{\hat{l}}},\acap
\dot{\mathit{\Omega}}^{Q_2} \lb{dotOQ2}& = -\rp{3}{2}\nk J_2\csc I\ton{\rp{R}{p}}^2\ton{\boldsymbol{\hat{k}}\boldsymbol\cdot\boldsymbol{\hat{h}}}\ton{\boldsymbol{\hat{k}}\boldsymbol\cdot\boldsymbol{\hat{m}}}.
\end{align}
In \rfrs{dotIQ2}{dotOQ2}, $\nk:=\sqrt{GM/a^3}$ is the Keplerian mean motion, $p:=a\ton{1-e^2}$ is the semilatus rectum, the unit vector $\boldsymbol{\hat{h}}$, directed along the orbital angular momentum and orthogonal to both $\boldsymbol{\hat{l}}$ and $\boldsymbol{\hat{m}}$ in such a way that $\boldsymbol{\hat{l}}\boldsymbol\times\boldsymbol{\hat{m}} = \boldsymbol{\hat{h}}$ holds, is defined as \citep{Sof89,1991ercm.book.....B,SoffelHan19,2024gpno.book.....I}
\eqi
\boldsymbol{\hat{h}} \lb{acca} := \grf{\sin I \sin\mathit{\Omega}, -\sin I \cos\mathit{\Omega}, \cos I},
\eqf
$R$ is the equatorial radius of the primary, and $J_2$ is its dimensionless mass quadrupole moment.
The latter one must be expressed in terms of its dimensional counterpart $Q_2$ having dimensions of a mass times a length squared as
\eqi
J_2 := -\rp{Q_2}{M R^2},
\eqf
which, for a Kerr BH, is given by \rfr{Q2}.

The resulting NH orbital rates are
\begin{align}
\dot I^\mathrm{NH} \lb{dotI}&= \dot I^\mathrm{LT} + \dot I^{Q_2},\acap
\dot{\mathit{\Omega}}^\mathrm{NH} \lb{dotO}&= \dot{\mathit{\Omega}}^\mathrm{LT} + \dot{\mathit{\Omega}}^{Q_2}.
\end{align}

\rfrs{dotILT}{acca} can be used to express the precession of the unit vector $\boldsymbol{\hat{h}}$ of the orbital angular momentum in a compact form as
\eqi
\dert{\boldsymbol{\hat{h}}}{t} \lb{vecprec} = \derp{\boldsymbol{\hat{h}}}{I}\,\dot I^\mathrm{NH} + \derp{\boldsymbol{\hat{h}}}{\mathit{\Omega}}\,\dot {\mathit{\Omega}}^\mathrm{NH} = {\boldsymbol{\Omega}}_\mathrm{d}^\mathrm{NH}\boldsymbol\times\boldsymbol{\hat{h}}.
\eqf
In it, the precession velocity vector ${\boldsymbol{\Omega}}^\mathrm{NH}_\mathrm{d}$ is
\eqi
{\boldsymbol{\Omega}}_\mathrm{d}^\mathrm{NH} = {\boldsymbol{\Omega}}_\mathrm{d}^\mathrm{LT} + {\boldsymbol{\Omega}}_\mathrm{d}^{Q_2},\lb{precNH}
\eqf
where
\begin{align}
{\boldsymbol{\Omega}}_\mathrm{d}^\mathrm{LT}  \lb{precLT} &=  \rp{2GJ}{c^2 a^3\ton{1-e^2}^{3/2}}\boldsymbol{\hat{k}},\acap
{\boldsymbol{\Omega}}_\mathrm{d}^{Q_2}  \lb{precQ2} &=  -\rp{3}{2}\nk J_2\ton{\rp{R}{p}}^2\ton{\boldsymbol{\hat{k}}\boldsymbol\cdot\boldsymbol{\hat{h}}}\boldsymbol{\hat{k}}.
\end{align}
From \rfrs{precNH}{precQ2}, it can be noted that the orbital angular momentum precesses around the primary's spin axis. Such results are in agreement with \citep{1974PhRvD..10.1340B,1975PhRvD..12..329B}, where  the term accounting also for the simultaneous precession of the Laplace--Runge--Lenz vector \citep{Gold80,Taff85}, of no interest here, was included in the total precessional velocity.
\section{Application to M87$^\ast$}\lb{sect:application}
In the following, the  spin axis of M87$^\ast$ will be parameterized as
\eqi
\boldsymbol{\hat{k}}=\grf{\sin\theta \sin\eta_\mathrm{p}, -\sin \theta \cos\eta_\mathrm{p}, \cos \theta},\lb{kappa}
\eqf
where \citep{2023Natur.621..711C}
\begin{align}
\theta \lb{theta}&= 17.21^\circ, \acap
\eta_\mathrm{p} \lb{etap}& = 288.47^\circ.
\end{align}

\begin{figure}
\centering
\begin{tabular}{c}
\includegraphics[width = 15 cm]{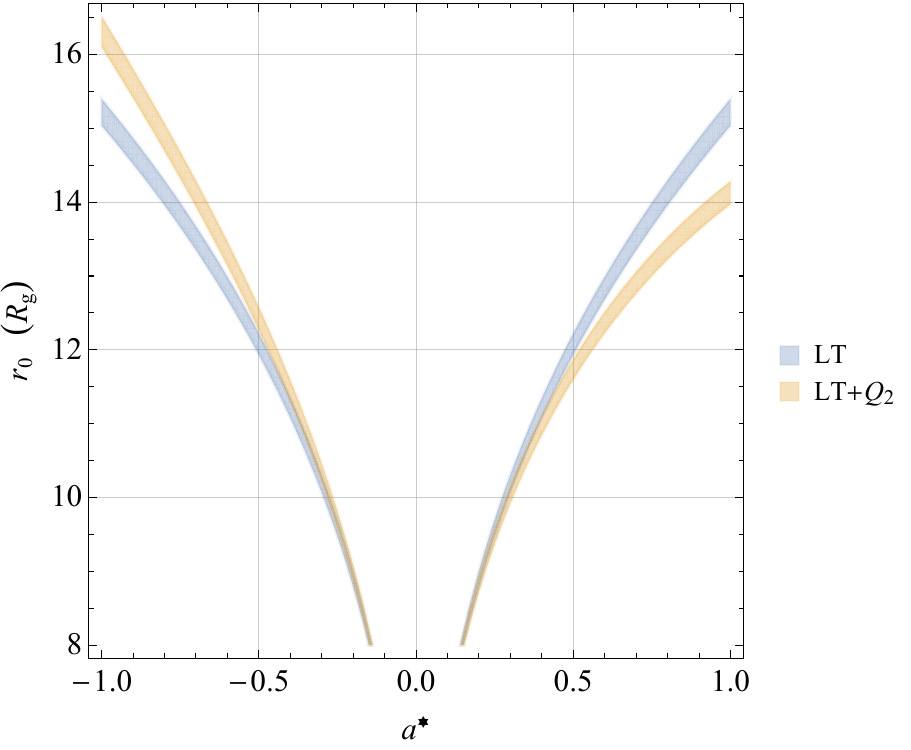}\\
\end{tabular}
\caption{Entire allowed regions in the $\grf{a^\ast,r_0}$ plane corresponding to the condition that the graphs of $\left|\Omega_\mathrm{d}^\mathrm{LT}\ton{a^\ast,r_0}\right|$ and $\left|\Omega_\mathrm{d}^\mathrm{NH}\ton{a^\ast,r_0}\right|$ remain confined between the upper and lower experimentally allowed values of $\left|\omega_\mathrm{p}^\mathrm{exp}\right|$, i.e., $0.54\,\mathrm{rad\,yr}^{-1}\leq \left|\Omega_\mathrm{d}^\mathrm{LT}\ton{a^\ast,r_0}\right|,\left|\Omega^\mathrm{NH}_\mathrm{d}\ton{a^\ast,r_0}\right|\leq 0.58\,\mathrm{rad\,yr}^{-1}$. The minimum value for $r_0$ has been taken from the largest expected radius of the TISCO orbit for $\psi_\mathrm{jet}\simeq 0^\circ$, as per Figure 1 of \citep{2024PhRvD.109b4029A}.
}\label{figure:allowed:tot}
\end{figure}

By considering the total NH precession velocity, given by \rfrs{precNH}{precQ2}, as a function of $a^\ast$ and the effective radius $r_0$ of the precessing accretion disk, it is possible to plot its absolute value by imposing the condition that its graph is comprised within the upper an lower measured values of the absolute value of the measured precession velocity   \citep{2023Natur.621..711C}
\eqi
\left|\omega^\mathrm{exp}_\mathrm{p}\right| = 0.56\pm 0.02\,\mathrm{rad\,yr}^{-1},
\eqf
i.e.,
\eqi
0.54\,\mathrm{rad\,yr}^{-1}\leq \left|\Omega_\mathrm{d}^\mathrm{NH}\ton{a^\ast,r_0}\right|\leq 0.58\,\mathrm{rad\,yr}^{-1}.\lb{condiz}
\eqf
The sets of the values of $a^\ast$ and $r_0$ satisfying the condition of \rfr{condiz} form allowed regions in the $\grf{a^\ast,r_0}$ plane depicted
in pale yellow in Figure \ref{figure:allowed:tot}.
In it, also the permitted regions obtained by considering solely the LT effect are shown in pale blue as well \citep{2024arXiv241108686I}. It turns out that both the LT and the $\mathrm{LT}+Q_2$ regions overlap in the no--rotation limit $\left|a^\ast\right|\rightarrow 0$, as expected. Instead, they separate when the hole's spin parameter tends to unity. While both the allowed branches are equal and symmetric with respect to the $a^\ast=0$ axis in the LT--only case, the inclusion of the SMBH's quadrupole breaks such a symmetry allowing for larger values of $r_0$ corresponding to the retrograde rotation of the hole ($a^\ast<0$) with respect to the prograde case ($a^\ast>0$).
As a general feature, for prograde hole's rotation, $Q_2$ generally tends to shrink the disk (right branches), while the opposite occurs for retrograde hole's rotation (left branches) with respect to the LT--only scenario. The minimum physically admissible value for $r_0$ was taken from Figure 1 of \citep{2024PhRvD.109b4029A} which depicts the radius of the tilted\footnote{The tilt is with respect to the hole's equator.} innermost stable circular orbit (TISCO) for different values of the spin--orbit tilt angle, called $\psi_\mathrm{jet}$ in  \citep{2023Natur.621..711C} (see below for more on it). It turns out that, for slightly tilted orbits as in the case of M87$^\ast$ ($\psi^\mathrm{M87^\ast}_\mathrm{jet}\simeq 1.25^\circ$ \citep{2023Natur.621..711C}), the radius of the TISCO is
\begin{align}
r_\mathrm{TISCO} &\simeq 8 R_\mathrm{g}\,\ton{a^\ast\simeq -1},\acap
r_\mathrm{TISCO} &\simeq 2 R_\mathrm{g}\,\ton{a^\ast\simeq 1}.
\end{align}

\begin{figure}
\centering
\begin{tabular}{c}
\includegraphics[width = 10 cm]{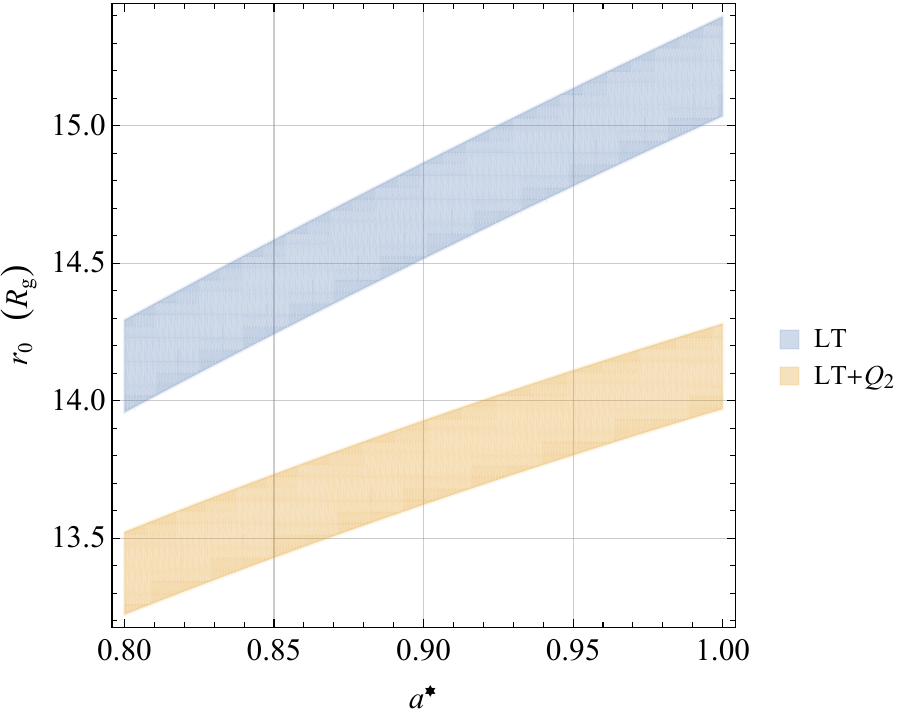}\\
\includegraphics[width = 10 cm]{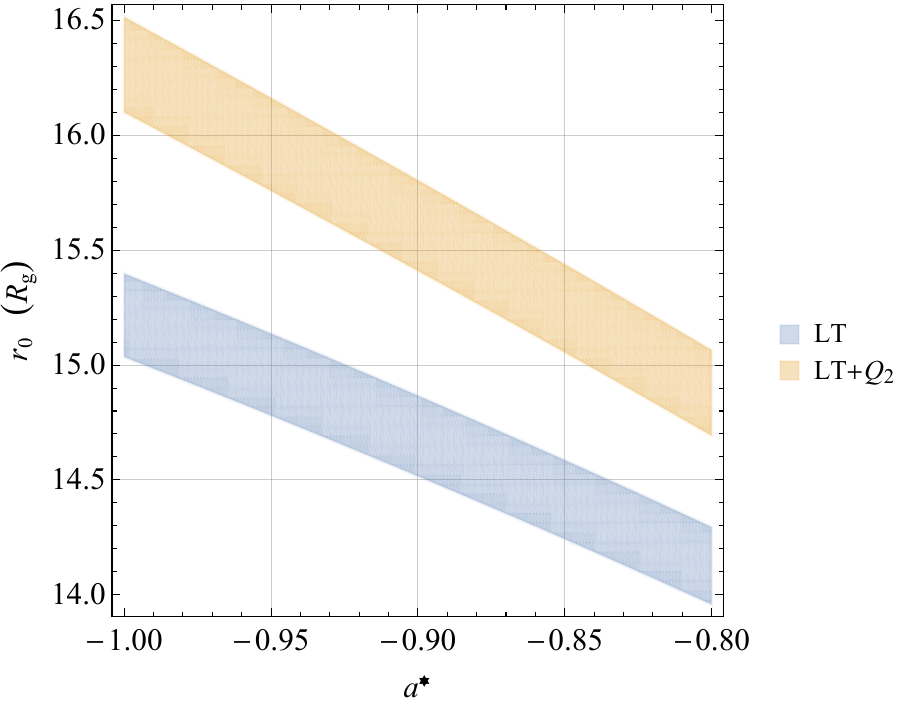}\\
\end{tabular}
\caption{Insets of  Figure \ref{figure:allowed:tot} for positive (upper panel) and negative (lower panel) values of the SMBH's spin parameter close to unity. The pale blue and pale yellow allowed regions correspond to the LT--only and $\mathrm{LT}+Q_2$ cases, respectively. }\label{figure:allowed_plus_minus}
\end{figure}

Figure \ref{figure:allowed_plus_minus} shows two insets of Figure \ref{figure:allowed:tot} corresponding to $\left|a^\ast\right|\lesssim 1$, in agreement with most of the constraints on it existing in the literature \citep{1998AJ....116.2237K,2008ApJ...676L.109W,2009ApJ...699..513L,2012Sci...338..355D,2017MNRAS.470..612F,2018MNRAS.479L..65S,2019ApJ...880L..26N,2019MNRAS.489.1197N,2020AnP...53200480G,2020MNRAS.492L..22T,2023Astro...2..141D}.
In such regimes, the impact of $Q_2$ appears quite evident with respect to the case in which only the LT effect is taken into account.

\textcolor{black}{N}umerically integrated NH time series of the time--dependent angles $\eta\ton{t}$ and $\phi\ton{t}$ determining the orientation of the disk's orbital angular momentum in space \textcolor{black}{can be produced}; in the language of celestial mechanics and astrodynamics, they correspond to the longitude of the ascending node $\mathit{\Omega}$ and the inclination $I$, respectively. Their signatures were theoretically obtained by simultaneously integrating the averaged equations for the rates of change of \rfrs{dotI}{dotO} by adopting the values $a^\ast = +0.98, r_0=14.1\,R_g$  and $a^\ast = -0.95, r_0=16\,R_g$, contained in the pale yellow regions of the upper and lower panels of Figure \ref{figure:allowed_plus_minus}, respectively. \textcolor{black}{The time spans of the integration have the same length as in Extended Data Figure 4 of \citep{2023Natur.621..711C} for a better comparison with the latter ones.}
\textcolor{black}{It turns out}  that the NH prograde signatures agree well with their measured counterparts of Figure 2 $\boldsymbol{\ton{\mathrm{b}}}$ and Extended Data Figure 4 of  \citep{2023Natur.621..711C}, both qualitatively and quantitatively. It is not the case for the NH retrograde signals  which, if on the one hand, retain the same sizes of those in Figure 2 $\boldsymbol{\ton{\mathrm{b}}}$ and Extended Data Figure 4 of  \citep{2023Natur.621..711C}, on the other hand, \textcolor{black}{are} out of phase with respect to the latter ones.

As far as the spin--orbit tilt angle $\psi_\mathrm{jet}$ is concerned, both its theoretically predicted value and its temporal evolution, calculated by integrating \rfrs{dotI}{dotO} as before, do not change with respect to their LT--only counterparts, shown in  Figure 5 of \citep{2024arXiv241108686I}, which were already found to be in agreement with their measured values.
\section{Summary and conclusions}\lb{fine}
The no--hair theorems are  fully at work in M87$^\ast$. Indeed, it has been shown that not only the hole's spin dipole moment $J$ through the Lense--Thirring effect but also its mass quadrupole moment $Q_2$, both calculated according to them, are able to explain the recently observed phenomenology of the jet precession of the supermassive black hole lurking at the centre of the galaxy M87. The inclusion of the precession induced by $Q_2$, calculated perturbatively as the Lense--Thirring one, yields to a modification of the allowed regions in the parameter space spanned by the hole's spin parameter $a^\ast$ and the effective radius $r_0$ of the accretion disk, assumed tightly coupled with the jet, with respect to the Lense--Thirring only case. Indeed, while neglecting $Q_2$ one obtains two identical permitted branches in the $\grf{a^\ast,r_0}$ plane which are symmetric with respect to the $a^\ast=0$ axis, the introduction of the hole's oblateness breaks such a symmetry since the allowed branch corresponding to retrograde hole's rotation gives larger allowed values for $r_0$ with respect to the prograde one. Furthermore, the presence of $Q_2$ tends to shrink (enlarge) the disk for prograde (retrograde) rotation with respect to the scenario in which it is omitted. The simultaneous numerical integration of the averaged equations for the spin dipole and mass quadrupole rates of change of the angles determining the orientation of the disk's orbital angular momentum with $a^\ast=0.98$ and $r_0=14.1$ gravitational radii allows to reproduce all the qualitative and quantitative features of the corresponding measured time series. It is not so for negative values of the hole's spin and the corresponding disk's radii which return out of phase theoretical signatures with respect to the empirical ones, while maintaining the same amplitudes. Finally, the value of the spin--orbit tilt angle between the angular momenta of the hole and the disk remains unchanged, along with its constancy over time.

\section*{Data availability}
No new data were generated or analysed in support of this research.
\section*{Conflict of interest statement}
I declare no conflicts of interest.
\section*{Acknowledgements}
I am grateful to M. Zaja\v{c}ek and Cui Y. for useful information and explanations. \textcolor{black}{I wish to thank also an anonymous referee for the appreciation paid to this work.}
\bibliography{Megabib}{}

\begin{thebibliography}{44}
\expandafter\ifx\csname natexlab\endcsname\relax\def\natexlab#1{#1}\fi

\bibitem[{{Al Zahrani}(2024)}]{2024PhRvD.109b4029A}
{Al Zahrani} A.~M., 2024, Phys. Rev. D, 109, 024029

\bibitem[{{Bardeen}(1970)}]{1970Natur.226...64B}
{Bardeen} J.~M., 1970, Nature, 226, 64

\bibitem[{{Barker} \& {Oconnell}(1974)}]{1974PhRvD..10.1340B}
{Barker} B.~M., {Oconnell} R.~F., 1974, Phys. Rev. D, 10, 1340

\bibitem[{{Barker} \& {O'Connell}(1975)}]{1975PhRvD..12..329B}
{Barker} B.~M., {O'Connell} R.~F., 1975, Phys. Rev. D, 12, 329

\bibitem[{{Bertotti}, {Farinella} \& {Vokrouhlick\'{y}}(2003){Bertotti},
  {Farinella}, \& {Vokrouhlick\'{y}}}]{2003ASSL..293.....B}
{Bertotti} B., {Farinella} P., {Vokrouhlick\'{y}} D., 2003, {Physics of the
  Solar System}. Kluwer

\bibitem[{{Brumberg}(1991)}]{1991ercm.book.....B}
{Brumberg} V.~A., 1991, {Essential Relativistic Celestial Mechanics}. Adam
  Hilger

\bibitem[{{Carter}(1971)}]{1971PhRvL..26..331C}
{Carter} B., 1971, Phys. Rev. Lett., 26, 331

\bibitem[{{Cui} {et~al}\mbox{.}(2023){Cui}, {Hada}, {Kawashima},
  {et~al.}}]{2023Natur.621..711C}
{Cui} Y., {Hada} K., {Kawashima} T., {et~al.}, 2023, Nature, 621, 711

\bibitem[{{Cui} \& {Lin}(2024)}]{2024arXiv241010965C}
{Cui} Y., {Lin} W., 2024, arXiv e-prints, arXiv:2410.10965

\bibitem[{{Doeleman} {et~al}\mbox{.}(2012){Doeleman}, {Fish}, {Schenck},
  {et~al.}}]{2012Sci...338..355D}
{Doeleman} S.~S., {Fish} V.~L., {Schenck} D.~E., {et~al.}, 2012, Science, 338,
  355

\bibitem[{{Dokuchaev}(2023)}]{2023Astro...2..141D}
{Dokuchaev} V.~I., 2023, Astronomy, 2, 141

\bibitem[{{Feng} \& {Wu}(2017)}]{2017MNRAS.470..612F}
{Feng} J., {Wu} Q., 2017, Mon. Not. Roy. Astron. Soc., 470, 612

\bibitem[{{Garofalo}(2020)}]{2020AnP...53200480G}
{Garofalo} D., 2020, Ann. Phys.--Berlin, 532, 1900480

\bibitem[{{Goldstein}(1980)}]{Gold80}
{Goldstein} H., 1980, {Classical Mechanics. Second Edition}. Addison Wesley

\bibitem[{{Iorio}(2024{\natexlab{a}})}]{2024gpno.book.....I}
{Iorio} L., 2024{\natexlab{a}}, {General Post-Newtonian Orbital Effects From
  Earth's Satellites to the Galactic Center}. Cambridge University Press

\bibitem[{{Iorio}(2024{\natexlab{b}})}]{2024arXiv241108686I}
{Iorio} L., 2024{\natexlab{b}}, arXiv e-prints, arXiv:2411.08686

\bibitem[{{Israel}(1967)}]{1967PhRv..164.1776I}
{Israel} W., 1967, Phys. Rev., 164, 1776

\bibitem[{{Kerr}(1963)}]{1963PhRvL..11..237K}
{Kerr} R.~P., 1963, Phys. Rev. Lett., 11, 237

\bibitem[{{Kissler-Patig} \& {Gebhardt}(1998)}]{1998AJ....116.2237K}
{Kissler-Patig} M., {Gebhardt} K., 1998, Astron J., 116, 2237

\bibitem[{{Kopeikin}, {Efroimsky} \& {Kaplan}(2011){Kopeikin}, {Efroimsky}, \&
  {Kaplan}}]{2011rcms.book.....K}
{Kopeikin} S.~M., {Efroimsky} M., {Kaplan} G., 2011, {Relativistic Celestial
  Mechanics of the Solar System}. Wiley

\bibitem[{{Li} {et~al}\mbox{.}(2009){Li}, {Yuan}, {Wang},
  {et~al.}}]{2009ApJ...699..513L}
{Li} Y.-R., {Yuan} Y.-F., {Wang} J.-M., {et~al.}, 2009, Astrophys. J., 699, 513

\bibitem[{{Liska} {et~al}\mbox{.}(2018){Liska}, {Hesp}, {Tchekhovskoy},
  {et~al.}}]{2018MNRAS.474L..81L}
{Liska} M., {Hesp} C., {Tchekhovskoy} A., {et~al.}, 2018, Mon. Not. Roy.
  Astron. Soc., 474, L81

\bibitem[{{McKinney}, {Tchekhovskoy} \& {Blandford}(2013){McKinney},
  {Tchekhovskoy}, \& {Blandford}}]{2013Sci...339...49M}
{McKinney} J., {Tchekhovskoy} A., {Blandford} R.~D., 2013, Science, 339, 49

\bibitem[{{Meng}, {Wang} \& {Wei}(2024){Meng}, {Wang}, \&
  {Wei}}]{2024arXiv241107481M}
{Meng} X.-C., {Wang} C.-H., {Wei} S.-W., 2024, arXiv e-prints, arXiv:2411.07481

\bibitem[{{Murray} \& {Dermott}(1999)}]{2000ssd..book.....M}
{Murray} C.~D., {Dermott} S.~F., 1999, {Solar System Dynamics}. Cambridge
  University Press

\bibitem[{{Nemmen}(2019)}]{2019ApJ...880L..26N}
{Nemmen} R., 2019, Astrophys. J. Lett., 880, L26

\bibitem[{{Newman} {et~al}\mbox{.}(1965){Newman}, {Couch}, {Chinnapared},
  {et~al.}}]{1965JMP.....6..918N}
{Newman} E.~T., {Couch} E., {Chinnapared} K., {et~al.}, 1965, J. Math. Phys.,
  6, 918

\bibitem[{{Newman} \& {Janis}(1965)}]{1965JMP.....6..915N}
{Newman} E.~T., {Janis} A.~I., 1965, J. Math. Phys., 6, 915

\bibitem[{{Nokhrina} {et~al}\mbox{.}(2019){Nokhrina}, {Gurvits}, {Beskin},
  {et~al.}}]{2019MNRAS.489.1197N}
{Nokhrina} E.~E., {Gurvits} L.~I., {Beskin} V.~S., {et~al.}, 2019, Mon. Not.
  Roy. Astron. Soc., 489, 1197

\bibitem[{{Penrose}(1999)}]{1999JApA...20..233P}
{Penrose} R., 1999, J. Astrophys. Astron., 20, 233

\bibitem[{{Penrose}(2002)}]{2002GReGr..34.1141P}
{Penrose} R., 2002, Gen. Relativ. Gravit., 7, 1141

\bibitem[{{Poisson} \& {Will}(2014)}]{2014grav.book.....P}
{Poisson} E., {Will} C.~M., 2014, {Gravity. Newtonian, Post--Newtonian,
  Relativistic}. Cambridge University Press

\bibitem[{{Robinson}(1975)}]{1975PhRvL..34..905R}
{Robinson} D.~C., 1975, Phys. Rev. Lett., 34, 905

\bibitem[{{Roy}(2005)}]{2005ormo.book.....R}
{Roy} A.~E., 2005, {Orbital Motion. Fourth Edition}. IOP Publishing

\bibitem[{{Shapiro} \& {Teukolsky}(1986)}]{1986bhwd.book.....S}
{Shapiro} S.~L., {Teukolsky} S.~A., 1986, {Black Holes, White Dwarfs and
  Neutron Stars: The Physics of Compact Objects}. Wiley

\bibitem[{{Shapiro} \& {Teukolsky}(1991)}]{1991PhRvL..66..994S}
{Shapiro} S.~L., {Teukolsky} S.~A., 1991, Phys. Rev. Lett., 66, 994

\bibitem[{{Sob'yanin}(2018)}]{2018MNRAS.479L..65S}
{Sob'yanin} D.~N., 2018, Mon. Not. Roy. Astron. Soc., 479, L65

\bibitem[{{Soffel}(1989)}]{Sof89}
{Soffel} M.~H., 1989, Relativity in Astrometry, Celestial Mechanics and
  Geodesy. Springer

\bibitem[{{Soffel} \& {Han}(2019)}]{SoffelHan19}
{Soffel} M.~H., {Han} W.-B., 2019, {Applied General Relativity}, {Astronomy and
  Astrophysics Library}. Springer

\bibitem[{{Taff}(1985)}]{Taff85}
{Taff} L.~G., 1985, {Celestial Mechanics: A Computational Guide for the
  Practitioner}. Wiley

\bibitem[{{Tamburini}, {Thid{\'e}} \& {Della Valle}(2020){Tamburini},
  {Thid{\'e}}, \& {Della Valle}}]{2020MNRAS.492L..22T}
{Tamburini} F., {Thid{\'e}} B., {Della Valle} M., 2020, Mon. Not. Roy. Astron.
  Soc., 492, L22

\bibitem[{{Teukolsky}(2015)}]{2015CQGra..32l4006T}
{Teukolsky} S.~A., 2015, Class. Quantum Gravit., 32, 124006

\bibitem[{{Wang} {et~al}\mbox{.}(2008){Wang}, {Li}, {Wang}, \&
  {Zhang}}]{2008ApJ...676L.109W}
{Wang} J.-M., {Li} Y.-R., {Wang} J.-C., {Zhang} S., 2008, Astrophys. J. Lett.,
  676, L109

\bibitem[{{Yodzis}, {Seifert} \& {M{\"u}ller Zum Hagen}(1973){Yodzis},
  {Seifert}, \& {M{\"u}ller Zum Hagen}}]{1973CMaPh..34..135Y}
{Yodzis} P., {Seifert} H.-J., {M{\"u}ller Zum Hagen} H., 1973, Communications
  in Mathematical Physics, 34, 135

\end{thebibliography}
\end{document}